\documentclass[]{llncs}

\usepackage[utf8]{inputenc}
\usepackage[T1]{fontenc} 
\usepackage[english]{babel}
\usepackage{amsmath,amssymb,latexsym,eufrak,euscript}
\usepackage{subfigure,pstricks,pst-node,pst-coil}

\usepackage{url,tikz}

\usepackage{multicol} 
\usepackage{multirow} 

\usepackage{bigcenter} 

\definecolor{darkblue}{rgb}{0,0,.3}					
\PassOptionsToPackage{bookmarksdepth=2}{hyperref}	
\usepackage[pdftex, colorlinks=true, linkcolor=darkblue,anchorcolor=darkblue, citecolor=darkblue, filecolor=darkblue, menucolor=darkblue, runcolor=darkblue, urlcolor=darkblue]{hyperref}			

\usetikzlibrary{arrows,snakes,automata,shapes}

\def \sp {\hspace*{0.8cm}}
\def \spt {\hspace*{0.3cm}}
\def \A {\mathcal{A}}
\def \T {\mathcal{T}}
\def \R {\mathcal{R}}
\def \B {\mathcal{B}}
\def \C {\mathcal{C}}

\def \N {\mathbb{N}}

\begin{document}

\label{firstpage}

  \title{Enhancing Approximations for Regular Reachability Analysis} 

\author{Alo\"is Dreyfus \and Pierre-Cyrille H\'eam \and Olga Kouchnarenko} 
\institute{FEMTO-ST CNRS 6174, University of Franche-Comt\'e \& Inria/CASSIS, France\\
\email{firstname.name@femto-st.fr}
}
 
\maketitle
\begin{abstract}
  This paper introduces two
  mechanisms for computing over-approximations of sets of reachable states, 
  with the aim of ensuring termination of state-space exploration. 
  The first mechanism consists in over-approximating the automata
  representing reachable sets by merging some of their states 
  with respect to simple syntactic criteria, or a combination of such criteria. 
  The second approximation mechanism consists in manipulating 
  an auxiliary automaton when applying a transducer
  representing the transition relation to an automaton encoding the initial states.  
  In addition, for the second mechanism we propose
  a new approach to refine the approximations depending on a property of interest. 
  The proposals are evaluated on examples of mutual
  exclusion protocols.
\end{abstract}

\section{Introduction and Problem Statement}

Reachability analysis is a challenging issue in formal software verification. 
Since the reachability problem is in general undecidable in most
formalisms, several ad-hoc approaches have been developed, such as symbolic 
reachability analysis using finite representations of infinite sets of
states.  {\it Regular model checking} (RMC for short) -- a symbolic approach using regular sets to represent sets
of states -- tackles undecidability in either of two ways: pointing out classes of regulars sets and relations for which the
reachability problem is decidable (see for
instance~\cite{DBLP:conf/icalp/GomezGP08}), or developing semi-algorithmic and/or approximation-based
approaches (see for instance~\cite{dams_iterating_2001,dams_iterating_2002}) to
semi-decide the reachability problem.  

In this paper we present new approximation techniques for RMC,  with the aim of 
providing quite efficient (semi-)algorithms.  The first technique consists in over-approximating the automata
  representing reachable sets by merging some of their states with respect to simple syntactic criteria, or a combination of such criteria (Section~\ref{quotient-based-approximations}).  
The second approximation technique consists in using  an auxiliary automaton when applying a transducer
  representing the transition relation to an automaton encoding the initial states (Section~\ref{transducer-based-approximations}). Moreover, for the second technique we develop  a new approach to refine the approximations, close to 
the well-known CEGAR technique (Section~\ref{transducer-based-approximations-CEGAR}).  The proposals are evaluated on examples of mutual
  exclusion protocols (Section~\ref{sec:experimentations}).
Omitted proofs are available online\footnote{\url{http://disc.univ-fcomte.fr/~adreyfus/ciaa13/version_longue.pdf}}.

\paragraph*{Related Work.}\label{sec:relatedwork}
Regular model-checking remains an active research domain in
computer science (see \cite{clarke_model_2000} and
\cite{baier_principles_2008} for a thorough overview).
In~\cite{kesten_symbolic_1997} the authors propose to use regular sets of
strings to represent states of parametrized arrays of processes, and to
represent the effect of performing an action by a predicate transformer
(transducer). In this work only transducers representing the effect of a
single application of a transition are considered, and consequently the
reachability analysis does not terminate for a lot of protocols. To bypass
this problem and still reach a fixpoint, the principal methods are
acceleration (providing exact computations)
\cite{jonsson_transitive_2000,bouajjani_regular_2000,dams_iterating_2001,dams_iterating_2002,abdulla_algorithmic_2003,boigelot_iterating_2003},
widening (extrapolating)
\cite{bouajjani_regular_2000,touili_regular_2001,legay_extrapolating_2012},
and automata abstraction \cite{bouajjani_abstract_2004}. Recently, new
results in RMC have been obtained for specific protocols (i.e.,
CLP~\cite{fioravanti_program_2011}, communicating
systems~\cite{Abstract_interpretation_of_FIFO_channels}, tree
language~\cite{abdulla_regular_tree_2002,bouajjani_widening_2011}, or
relational string verification using multi-track
automata~\cite{yu_relational_2011}), using domain-specific
techniques~\cite{boigelot_domain-specific_2012}. Our contributions aim at 
improving the generic method in~\cite{bouajjani_abstract_2004} by giving means
to build over-approximations by merging abstract states of the system
(and not of the transducer, which is never modified).
Unlike~\cite{bouajjani_regular_2000,bouajjani_abstract_2004}, our proposals 
do not require the subset-construction, minimization and 
determinization of the obtained automaton at each RMC step. 

\begin{figure}[tb]
\centering
\begin{minipage}{0.3\textwidth}
\centering
 \subfigure[$\A_1$]{
 \begin{tikzpicture}[scale=0.6, every node/.style={scale=0.9}]
 \node (1)[state,draw,fill=blue!20,initial, initial text=] at (0,0) {$1$};
 \node (2)[state,draw,fill=blue!20,initial, initial text=,accepting,] at (3,0)
       {$2$};
 \path[->,>=triangle 90] (1) edge[out=45] node [above] {$a$} (2);
 \path[->,>=triangle 90] (1) edge[loop left] node [above] {$a,b$} ();
 \path[->,>=triangle 90] (2) edge[loop right] node [above] {$b$} ();
 \end{tikzpicture}
 }
\subfigure[$\mathcal{T}$]{
\begin{tikzpicture}[scale=0.6, every node/.style={scale=0.9}]
\node (1)[state,draw,fill=blue!20,initial, initial text=] at (0,0) {$4$};
\node (2)[state,draw,fill=blue!20,accepting,] at (3,0)
      {$3$};
\path[->,>=triangle 90] (1) edge[out=45] node [above] {$(a,b)$} (2);
\path[->,>=triangle 90] (1) edge[loop above] node [above] {$(b,a)$} ();
\path[->,>=triangle 90] (2) edge[out=-125,in=-45] node [above] {$(b,a)$} (1);
\path[->,>=triangle 90] (2) edge[loop above] node [above] {$(a,b)$} ();
\end{tikzpicture}
}
\end{minipage}
\begin{minipage}{0.3\textwidth}
\centering
\subfigure[$\mathcal{T}(\A_1)$]{
\begin{tikzpicture}[scale=0.7]
\node (14)[state,draw,fill=blue!20,initial, initial text=] at (0,0) {$1,4$};
\node (24)[state,draw,fill=blue!20,initial, initial text=] at (3,0)
      {$2,4$};
\node (13)[state,draw,fill=blue!20] at (0,-3) {$1,3$};
\node (23)[state,draw,fill=blue!20,accepting,] at (3,-3) {$2,3$};
\path[->,>=triangle 90] (14) edge[] node [above] {$b$} (23);
\path[->,>=triangle 90] (14) edge[] node [left] {$b$} (13);
\path[->,>=triangle 90] (23) edge[] node [left] {$a$} (24);
\path[->,>=triangle 90] (13) edge[] node [below] {$b$} (23);
\path[->,>=triangle 90] (13) edge[out=45,in=-60] node [right] {$a$} (14);

\path[->,>=triangle 90] (14) edge[loop above] node [above] {$a$} ();
\path[->,>=triangle 90] (24) edge[loop above] node [above] {$a$} ();
\path[->,>=triangle 90] (13) edge[loop left] node [] {$b$} ();
\end{tikzpicture}
}
\end{minipage}
\begin{minipage}{0.3\textwidth}
\centering
\subfigure[$\mathcal{T}(\A_1)/_{\sim_{\rm exe}}$]{
\begin{tikzpicture}[scale=0.7]
\node (1)[state,draw,fill=blue!20,initial, initial text=] at (0,0) {};
\node (2)[state,draw,fill=blue!20,accepting] at (3,0) {2,3};
\path[->,>=triangle 90] (1) edge[out=45] node [above] {$b$} (2);
\path[->,>=triangle 90] (2) edge[out=225,in=-45] node [above] {$a$} (1);
\path[->,>=triangle 90] (1) edge[loop above] node [above] {$a,b$} ();
\end{tikzpicture}
}
\end{minipage}
\vspace{-1em}
\caption{Illustrating examples}\label{fig:def}
\end{figure}
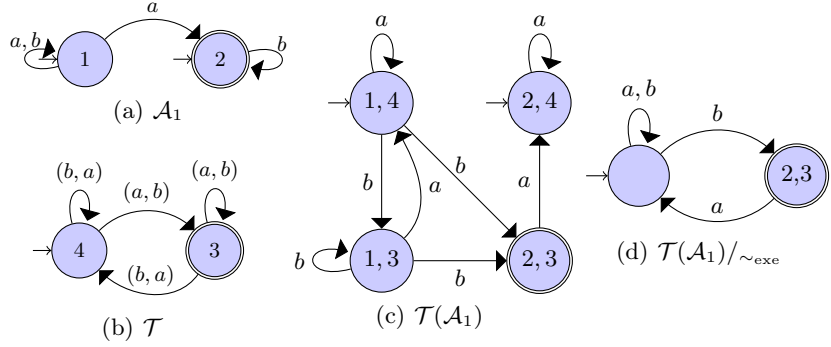

\paragraph*{Formal Background.}
We assume the reader familiar with basic notions of language theory.
%
An {\it automaton} $\A$ on an alphabet $\Sigma$ is a tuple
$(Q,\Sigma,E,I,F)$ where $Q$ is the finite set of {\it states}, $E\subseteq
Q\times \Sigma\times Q$ is the set of {\it transitions}, $I\subseteq Q$ is
the set of {\it initial states} and $F\subseteq Q$ is the set of {\it final
  states}. We define the size of $\A$ by $|\A| = |Q| + |E|$. An automaton is
{\it deterministic} [resp. {\it complete}] if $I$ is a singleton and for
each $(q,a)\in Q\times \Sigma$ there is at most [resp. at least] one $p\in
Q$ such that $(q,a,p)\in E$. A path in $\A$ is a (possibly empty) finite
sequence of transitions $(p_1,a_1,q_1)\ldots (p_n,a_n,q_n)$ such that for
each $i$, $q_i=p_{i+1}$. The integer $n$ is the length of the path and the
word $a_1\ldots a_n$ is its label. A path is {\it successful} if $p_1$ is
initial and $p_n$ is final. A word $w$ is {\it accepted} by $\A$ if $w$ is
the label of a successful path. The set of words accepted by $\A$ is denoted
$L(\A)$. If $\A$ is deterministic and complete, for every state $q$ and
every word $w$, there exists a unique state of $\A$, denoted $q\cdot_\A w$
reachable from $q$ by reading a path labeled by $w$. If there is no
ambiguity on $\A$, it is simply denoted $q\cdot w$. By convention, $q \cdot
\varepsilon=\{q\}$. A state $q$ is {\it accessible} [resp. {\it
    co-accessible}] if there exists a path from an initial state to $q$
                 [resp. if there exists a path from $q$ to a final state].
                 An automaton whose states are all both accessible and
                 co-accessible is called {\it trim}. If $\A$ is not a trim
                 automaton, removing from $\A$ all states that are not both
                 accessible and co-accessible together with all related
                 transitions provides an equivalent trim automaton. Let
                 $\A_1=(Q_1,\Sigma,E_1,I_1,F_1)$ and
                 $\A_2=(Q_2,\Sigma,E_2,I_2,F_2)$ be two automata over the
                 same alphabet, the product of $\A_1$ and $\A_2$ is the
                 automaton $(Q_1\times Q_2,\Sigma,E,I_1\times I_2,F_1\times
                 F_2)$, denoted $\A_1\times \A_2$, where
                 $E=\{((p_1,p_2),a,(q_1,q_2))\mid (p_1,a,q_1)\in E_1 \wedge
                 (p_2,a,q_2)\in E_2\}$. By definition, $L(\A_1\times
                 \A_2)=L(\A_1)\cap L(\A_2)$. 
                 Let
                 $\hat{\A}=(\hat{Q},\Sigma,\hat{E},\hat{I},\hat{F})$
                 be the trim automaton obtained from $\A$, given an
                 equivalence relation $\sim \subseteq Q\times Q$,
                 $\A/_\sim$ denotes the automaton
                 $(\hat{Q}/_\sim,\Sigma,E^\prime,\hat{I}/_\sim,
                 \hat{F}/_\sim)$ where
                 $E^\prime=\{(\tilde{p},a,\tilde{q})\mid \exists p\in
                 \tilde{p}\text{ and } \exists q\in \tilde{q}\text{
                   s.t. } (p,a,q)\in \hat{E}\}$. One can easily check
                 that $L(\A)\subseteq L(\A/_\sim)$. For instance,
                 given the automata of Fig.~\ref{fig:def} and the
                 relation $\sim_{\rm exe}$ whose classes are $\{(1,4),
                 (2,4), (1,3)\}$ and $\{(2,3)\}$, the automaton
                 $\mathcal{T}(\A_1)/_{\sim_{\rm exe}}$ is depicted on
                 Fig.~\ref{fig:def}. Two automata
                 $\A_1=(Q_1,\Sigma,E_1,I_1,F_1)$ and
                 $\A_2=(Q_2,\Sigma,E_2,I_2,F_2)$ are isomorphic if
                 there exists a one-to-one function $f\,:\, Q_1
                 \rightarrow Q_2$ satisfying $(p,a,q)\in E$ iff
                 $((f(p),a,f(q))\in E$, and $f(I_1)=I_2$, $f(F_1)=F_2$
                 when lifted to sets. Informally, two automata are
                 isomorphic if they are equal up to state names.

Let $\Sigma_1$ and $\Sigma_2$ be two alphabets, a {\it
  transducer} on $\Sigma_1,\Sigma_2$ is an automaton on
$\Sigma_1\times\Sigma_2$. Each transducer $\T$ on $\Sigma_1,\Sigma_2$
induces a relation $R_\T$ on $\Sigma_1^*\times\Sigma_2^*$ defined by:
for the $a_i$'s in $\Sigma_1$ and the $b_j$'s in $\Sigma_2$,  
$(a_1\ldots a_n,b_1\ldots b_m)\in R_\T$ iff $n=m$ and the word
$(a_1,b_1)\ldots(a_n,b_n)$ is accepted by $\T$.
The reflexive transitive closure of $R_\T$ is denoted $R_\T^*$.
Let $\A=(Q_1,\Sigma,E_1,I_1,F_1)$ be an automaton on $\Sigma_1$, and 
$\T=(Q_2,\Sigma_1\times \Sigma_2,E_2,I_2,F_2)$ a transducer on
$\Sigma_1 \times \Sigma_2$, we denote by $\T(\A)$ the automaton $(Q_1\times
Q_2,\Sigma_2,E,I_1\times I_2,F_1\times F_2)$ on $\Sigma_2$ where
$E=\{((p_1,p_2),b,(q_1,q_2))\mid (p_1,a,q_1)\in E_1 $ $\wedge 
(p_2,(a,b),q_2)$ $\in E_2\}$. 
An example is depicted on Fig.~\ref{fig:def}.
By definition, $L(\T(\A))$ is the set of words
$v$ satisfying $(u,v)\in R_\T$ for some words $u\in L(\A)$. 
If $\T=(Q_2,\Sigma_1\times \Sigma_2,E_2,I_2,F_2)$ is a transducer, we denote
by $\T^{-1}$ the transducer $(Q_2,\Sigma_2\times \Sigma_2,E_2^\prime,I_2,F_2)$
with $E^\prime_2=\{(p,(a,b),q)\mid (p,(b,a),q)\in E_2\}$. One can check that
$(u,v)\in R_\T$ iff $(v,u)\in R_{\T^{-1}}$.

\paragraph*{Regular Reachability Problem.}
The following regular reachability problem -- central for RMC -- is known to be undecidable in general; its variants have
been addressed in most of the papers in Sect.~\ref{sec:relatedwork}. \smallskip

\noindent
{\bf Input:} Two finite automata $\A$ and $\mathcal{B}$ on a same alphabet $\Sigma$, 
and a transducer $\T$ on $\Sigma\times\Sigma$.\\
{\bf Output:} {\bf 1} if $R_\T^*(L(\A))\cap L(\B)=\emptyset$, and {\bf 0}
otherwise. 
\smallskip


Since the problem is concerned with the reflexive-transitive closure, we may
assume without loss of generality that for every $u\in \Sigma^*$,
$(u,u)\in R_\T$. In the rest of the paper, all considered relations contain the
identity.

\section{Quotient-based Approximations}
\label{quotient-based-approximations}

This section introduces the first mechanism for computing
over-approximations of sets of reachable states, which consists in
over-approximating the automata representing reachable sets by merging
some of their states. For doing this, basic elementary policies as
well as their combinations are introduced.
  
Given an automaton $\A$, we define an {\it approximation} as a
function mapping each automaton $\A$ to an equivalence relation
$\sim_\A$ over the states of $\A$. The approximation function
$\mathfrak{F}$ is {\it isomorphism-compatible} if for every pair of
automata $\A_1$ and $\A_2$, every isomorphism $\varphi$ from $\A_1$ to $\A_2$,
$p\sim_{\A_1} q$ iff $\varphi(p)\sim_{\A_2} \varphi(q)$. We denote
$\mathfrak{F}[\A]$ the automaton $\hat{\A}/_{\mathfrak{F}(\hat{\A})}$,
where $\hat{\A}$ is the trim automaton obtained from $\A$. We
inductively define $\mathfrak{F}^n[\A]$ by $\mathfrak{F}^0[\A]=\A$, and
$\mathfrak{F}^{n}[\A]=\mathfrak{F}[\mathfrak{F}^{n-1}[\A]]$ for $n>0$.

Let us now introduce two isomorphism-compatible approximation
functions. They are easily computable, and represent simple criteria
naturally used by the specifier, as for example 
in~\cite{bouajjani_abstract_2004} for computing equivalence relations, or 
in~\cite{DBLP:conf/fm/BauerF12} for monitoring LTL properties.
\begin{itemize}
\item $\mathfrak{Left}$, mapping each automaton $(Q,\Sigma,E,I,F)$ to the reflexive-transitive \linebreak[4]
closure of the relation $R_{\rm left}$,
  defined by $p R_{\rm left} q$  iff $L(Q,\Sigma,E,I,\{p\}) \cap \linebreak[4] L(Q,\Sigma, E,I,\{q\}) \neq \emptyset$.

\item $\mathfrak{Right}$, mapping each automaton $(Q,\Sigma,E,I,F)$ to the
  reflexive-transitive closure of the relation $R_{\rm right}$, defined by
  $p R_{\rm right} q$ iff $L(Q,\Sigma,E,\{p\},F)\cap
  L(Q,\Sigma,E,\{q\},F)\neq \emptyset$.


\end{itemize}

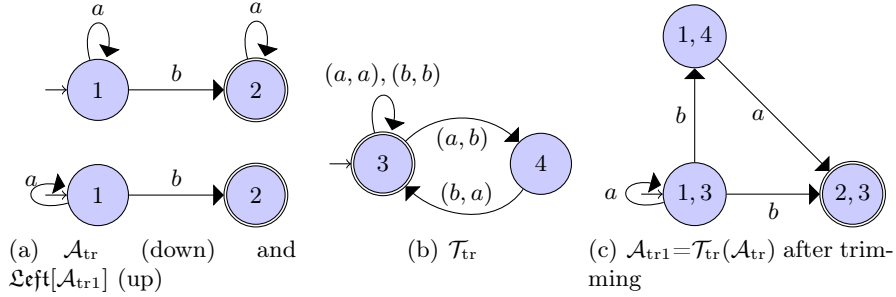
\begin{figure}[tb]
\subfigure[$\A_{\rm tr}$ (down) and $\mathfrak{Left}\lbrack\A_{\rm
    tr1}\rbrack$ (up)]{\label{fig:tr-Ta}
\begin{tikzpicture}[scale=0.7]

\node (1a)[state,draw,fill=blue!20,initial, initial text=] at (0,2) {1};
\node (2a)[state,draw,fill=blue!20,accepting] at (3,2)
      {2};

\path[->,>=triangle 90] (1a) edge[] node [above] {$b$} (2a);
\path[->,>=triangle 90] (1a) edge[loop above] node [above] {$a$} ();
\path[->,>=triangle 90] (2a) edge[loop above] node [above] {$a$} ();

\node (1)[state,draw,fill=blue!20,initial, initial text=] at (0,0) {1};
\node (2)[state,draw,fill=blue!20,accepting] at (3,0)
      {2};

\path[->,>=triangle 90] (1) edge[] node [above] {$b$} (2);
\path[->,>=triangle 90] (1) edge[loop left] node [above] {$a$} ();
\end{tikzpicture}
}
\subfigure[$\mathcal{T}_{\rm tr}$]{\label{fig:tr-T}
\begin{tikzpicture}[scale=0.7]
\node (1)[state,draw,fill=blue!20,initial, initial text=,accepting] at (0,0) {$3$};
\node (2)[state,draw,fill=blue!20] at (3,0)
      {$4$};

\path[->,>=triangle 90] (1) edge[out=45] node [below] {$(a,b)$} (2);
\path[->,>=triangle 90] (1) edge[loop above] node [above] {$(a,a),(b,b)$} ();
\path[->,>=triangle 90] (2) edge[out=-125,in=-45] node [above] {$(b,a)$} (1);

\end{tikzpicture}
}
\subfigure[$\A_{\rm tr1}$=$\mathcal{T}_{\rm tr}(\A_{\rm tr})$ after trimming]{
\begin{tikzpicture}[scale=0.7]
\node (14)[state,draw,fill=blue!20] at (0,0) {$1,4$};

\node (13)[state,draw,fill=blue!20,initial, initial text=] at (0,-3) {$1,3$};
\node (23)[state,draw,fill=blue!20,accepting,] at (3,-3)
      {$2,3$};

\path[->,>=triangle 90] (14) edge[] node [left] {$a$} (23);
\path[->,>=triangle 90] (13) edge[] node [left] {$b$} (14);
\path[->,>=triangle 90] (13) edge[] node [below] {$b$} (23);

\path[->,>=triangle 90] (13) edge[loop left] node [] {$a$} ();
\end{tikzpicture}
}

\caption{Token ring}\label{fig:tr1}
\end{figure}

Let us consider the example of the token ring protocol for which the
automata are depicted on Fig.~\ref{fig:tr1}. Let $\A_{\rm tr1}$ be the
automaton obtained by trimming $\mathcal{T}_{\rm tr}(\A_{\rm
  tr})$. The relation $\mathfrak{Right}[\A_{\rm tr1}]$ is the identity
relation, therefore $\mathfrak{Right}[\A_{\rm tr1}]=\A_{\rm
  tr1}$. However, for the relation $\mathfrak{Left}$, the states
$(1,4)$ and $(2,3)$ are equivalent since they can be reached from the
initial state by reading $b$. The automaton $\mathfrak{Left}[\A_{\rm
  tr1}]$ is depicted on Fig.~\ref{fig:tr-Ta} (up).

\begin{proposition}\label{prop-iso}
  For each automaton $\A$, if $\mathfrak{F}$ is an isomorphism-compatible approximation function, then the sequence
  $(\mathfrak{F}^n[\A])_{n\in\N}$ is ultimately constant, up to
  isomorphism. Let $C_\mathfrak{F}(\A)$ denote the limit of
  $(\mathfrak{F}^n[\A])_{n\in\N}$. Moreover, if for each automaton
  $\A$ and each pair of states $p,q$ of $\A$, one can check in
  polynomial time whether $p\sim_\A q$, then $C_\mathfrak{F}(\A)$ can
  be computed in polynomial time as well.
\end{proposition}



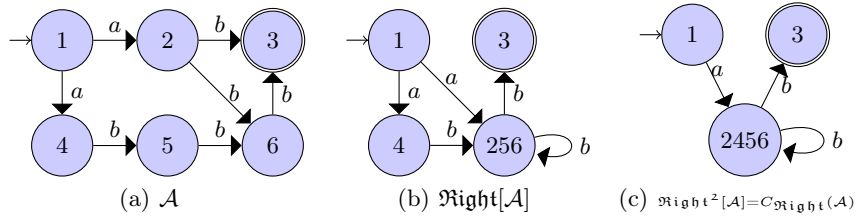
\begin{figure}[tb]
\subfigure[$\A$]{\label{fig:approx-A}
\begin{tikzpicture}[scale=0.7]
\node (1)[state,draw,fill=blue!20,initial, initial text=] at (0,1) {$1$};

\node (2)[state,draw,fill=blue!20,] at (2,1) {$2$};

\node (3)[state,draw,fill=blue!20,accepting] at (4,1) {$3$};

\node (4)[state,draw,fill=blue!20,] at (0,-1) {$4$};

\node (5)[state,draw,fill=blue!20,] at (2,-1) {$5$};

\node (6)[state,draw,fill=blue!20] at (4,-1) {$6$};

\path[->,>=triangle 90] (1) edge[] node [above] {$a$} (2);
\path[->,>=triangle 90] (1) edge[] node [right] {$a$} (4);

\path[->,>=triangle 90] (2) edge[] node [above] {$b$} (3);
\path[->,>=triangle 90] (2) edge[] node [right] {$b$} (6);

\path[->,>=triangle 90] (4) edge[] node [above] {$b$} (5);

\path[->,>=triangle 90] (5) edge[] node [above] {$b$} (6);

\path[->,>=triangle 90] (6) edge[] node [right] {$b$} (3);

\end{tikzpicture}
}
\subfigure[$\mathfrak{Right[\A]}$]{\label{fig:approx-Right-A}
\begin{tikzpicture}[scale=0.7]
\node (1)[state,draw,fill=blue!20,initial, initial text=] at (0,1) {$1$};

\node (3)[state,draw,fill=blue!20,accepting] at (2,1) {$3$};

\node (4)[state,draw,fill=blue!20,] at (0,-1) {$4$};

\node (256)[state,draw,fill=blue!20,] at (2,-1) {$256$};

\path[->,>=triangle 90] (1) edge[] node [above] {$a$} (256);
\path[->,>=triangle 90] (1) edge[] node [right] {$a$} (4);

\path[->,>=triangle 90] (256) edge[] node [right] {$b$} (3);
\path[->,>=triangle 90] (256) edge[loop right] node [right] {$b$} ();

\path[->,>=triangle 90] (4) edge[] node [above] {$b$} (256);

\end{tikzpicture}
}
\subfigure[{\tiny $\mathfrak{Right^2[\A]}$=$C_\mathfrak{Right}(\A)$}]{\label{fig:approx-Right2-A}
\begin{tikzpicture}[scale=0.7]
\node (1)[state,draw,fill=blue!20,initial, initial text=] at (0,1) {$1$};

\node (3)[state,draw,fill=blue!20,accepting] at (2,1) {$3$};

\node (2456)[state,draw,fill=blue!20,] at (1,-1) {$2456$};

\path[->,>=triangle 90] (1) edge[] node [above] {$a$} (2456);

\path[->,>=triangle 90] (2456) edge[] node [right] {$b$} (3);
\path[->,>=triangle 90] (2456) edge[loop right] node [right] {$b$} ();

\end{tikzpicture}
}
\caption{Computing $C_\mathfrak{Right}(\A)$}\label{fig:approx}
\end{figure}

\begin{figure}[tb]
\subfigure[{\tt FixPoint}]{\label{algo:fp}
\begin{tabular}{l}
{\bf Semi-Algorithm} {\tt FixPoint}\\
{\bf Input:} $\A$, $\T$, $\B$, $\mathfrak{F}$\\
\sp {\bf If} $L(C_\mathfrak{F} (\T(\A)))\cap L(\B)\neq \emptyset$ {\bf then}\\
\sp\sp {\bf return} {\it Inconclusive}\\
\sp {\bf EndIf}\\
\sp {\bf If} $L(C_\mathfrak{F} (\T(\A)))=L(\A)$  {\bf then}\\
\sp\sp {\bf return} {\it Safe}\\
\sp {\bf EndIf}\\
\sp {\bf Return} {\tt FixPoint}($C_\mathfrak{F}
(\T(\A)),\T$,$\B$,$\mathfrak{F}$)\\
 \\
\\
\\
\end{tabular}
}\subfigure[{\tt FixPointT}]{\label{algo:fpT}
\begin{tabular}{l}
{\bf Semi-Algorithm} {\tt FixPointT}\\
{\bf Input:} $\A$, $\T$, $\B$, $\C$\\
{\bf Variable:} $k$\\
\sp k:=0\\
\sp {\bf While} ($L(\T_\C^{k+1}(\A))\neq L(\T_\C^k(\A))$) {\bf do}\\
\sp\sp $k:=k+1$\\
\sp {\bf EndWhile}\\
\sp {\bf If} ($L(\T_\C^k(\A))\cap L(\B)= \emptyset$) {\bf then}\\
\sp\sp {\bf Return} {\it Safe}\\
\sp {\bf Else}\\
\sp\sp {\bf Return} {\it Inconclusive}\\
\sp {\bf EndIfElse}
\end{tabular}
}
\caption{Fixpoint algorithms}\label{algo:master}
\end{figure}

In the {\tt FixPoint} algorithm depicted in Fig. \ref{algo:fp},
given a finite automaton $\A$ (state of the system), a transducer $\T$ (transition relation), a finite automaton $\B$ (bad property), and an isomorphism-compatible function $\mathfrak{F}$ (approximation criterion), the first check (emptiness) can be performed in polynomial time. 
Then, unfortunately, the equality of the languages cannot
be checked in polynomial time, since the involved automata are not
deterministic. Nevertheless, recently developed
algorithms~\cite{DBLP:conf/tacas/DoyenR10,DBLP:conf/tacas/AbdullaCHMV10,BONCHI-2012-674660}
allow solving this problem very efficiently. Note also that the
equality test can be replaced by another test -- e.g., isomorphism or
(bi)simulation -- implying language equality or inclusion, as
$L(\A)\subseteq L(C_\mathfrak{F} (\T(\A)))$) by construction.

\begin{proposition}\label{prop-fixed}
The {\tt FixPoint}  semi-algorithm is correct: if it returns {\em Safe}, then $R_\T^*(L(\A))\cap L(\B)= \emptyset$.
\end{proposition}
 


The approach can be illustrated on the example in Fig.~\ref{fig:tr1} with
$\mathfrak{F}=\mathfrak{Left}$: $C_\mathfrak{Left}(\T_{\rm tr}(\A_{\rm
  tr}))=\mathfrak{Left}(\A_{\rm tr}).$ One can check that
 $C_\mathfrak{Left}(\T_{\rm tr}(C_\mathfrak{Left}(\T_{\rm tr}(\A_{\rm
  tr}))))$ and $C_\mathfrak{Left}(\T_{\rm tr}(\A_{\rm
  tr}))$ are isomorphic. Therefore {\tt FixPoint} stops after one
recursive call and returns {\it Safe}.


From now on, given two approximation functions $\mathfrak{F}$ and
$\mathfrak{G}$, we denote $\mathfrak{F}.\mathfrak{G}$ the
approximation function defined by
$(\mathfrak{F}.\mathfrak{G})(\A)=\mathfrak{F}(\A)\cap
\mathfrak{G}(\A)$ for every automaton $\A$. In addition, the
approximation function $\mathfrak{F}+\mathfrak{G}$ is defined by: for
every automaton $\A$, $(\mathfrak{F}+\mathfrak{G})(\A)$ is the
smallest equivalence relation containing both $\mathfrak{F}(\A)$ and
$\mathfrak{G}(\A)$. Then using several approximation functions and
combining them allow us to obtain new -- stronger or weaker --
approximations. Section~\ref{sec:experimentations}
gives experimental results for the $\mathfrak{Left}$, $\mathfrak{Right}$ approximations together with the $\mathfrak{In}$ and $\mathfrak{Out}$ approximations, and for their combinations.

\section{Transducer-based Approximations}
\label{transducer-based-approximations}

This section introduces another approximation mechanism consisting in
reasoning about the application of $k$ copies of a transducer
representing the transition relation to an automaton representing the initial states.  The states reached in the transducers are encoded as a finite word, and an additional automaton is used for specifying
what are the combinations of transducer states that have to be
merged. This technique is inspired by an automata theoretic construction in~\cite{bouajjani_regular_2000}, 
with the difference concerning the equivalence relation, and the use of automata at step $k$ (the transducer is never modified).

Let $\A=(Q,\Sigma,E,I,F)$ be a finite automaton,
$\mathcal{T}=(Q_T,\Sigma\times\Sigma,E_T,I_T,F_T)$ a transducer, and 
$\C=(Q_C,Q_T,E_C,\{q_{\rm init}\},\emptyset)$ a deterministic complete finite
automaton on $Q_T$ (i.e., the transitions of $\C$ are labeled with states of
$\mathcal{T}$).
%
Let $\varphi_k$ be a one-to-one mapping from the set
$(((Q\times Q_T)\times Q_T)\ldots \times Q_T)$ of states of $\T^k(\A)$ to
$Q\times Q_T^k$, where $Q_T^k$ is the set of words of length $k$ on $Q_T$. We
set a relation $\sim_\C$ on states of $\T^k(\A)$ as follows: if $p$ and $q$
are states of $\T^k(\A)$ such that $\varphi_k(p)=(p_0,w_p)$ and
$\varphi_k(q)=(q_0,w_q)$, then $p\sim_\C q$ iff $p_0=q_0$ and $q_{\rm
  init}\cdot w_p=q_{\rm init}\cdot w_q$. The automaton $\T^k(\A)/_{\sim_\C}$
is denoted $\T^k_\C(\A).$ One can easily check that $\sim_\C$ is an
equivalence relation.

\begin{figure}[tb]
\subfigure[$\mathcal{T}_{\rm tr}(\A_{\rm tr})/_{\sim_\C}$]{\label{fig:tr3-a}
\begin{tikzpicture}[scale=0.7]
\node (14)[state,draw,fill=blue!20] at (0,-1) {$1,\bigcirc$};

\node (13)[state,draw,fill=blue!20,initial, initial text=] at (0,-3) {$1,\square$};
\node (23)[state,draw,fill=blue!20,accepting,] at (3,-3)
      {$2,\square$};

\path[->,>=triangle 90] (14) edge[] node [left] {$a$} (23);
\path[->,>=triangle 90] (13) edge[] node [left] {$b$} (14);
\path[->,>=triangle 90] (13) edge[] node [below] {$b$} (23);

\path[->,>=triangle 90] (13) edge[loop left] node [] {$a$} ();
\end{tikzpicture}
}\subfigure[$\mathcal{T}_{\rm tr}^2(\A_{\rm tr})/_{\sim_C}$]{\label{fig:tr3-b}
\begin{tikzpicture}[scale=0.7]
\node (133)[state,draw,fill=blue!20,initial, initial text=] at (0,-1)
      {$1,q_{\rm init}$};

\node (233)[state,draw,fill=blue!20,accepting] at (2,1)
      {$2,q_{\rm init}$};

\node (134)[state,draw,fill=blue!20,] at (4,-1)
      {$1,\bigtriangledown$};

\path[->,>=triangle 90] (133) edge[loop left] node [left] {$a$} ();
\path[->,>=triangle 90] (134) edge[loop right] node [right] {$a$} ();
\path[->,>=triangle 90] (133) edge[] node [above] {$b$} (134);
\path[->,>=triangle 90] (133) edge[] node [below] {$b$} (233);
\path[->,>=triangle 90] (134) edge[] node [left] {$a$} (233);
\end{tikzpicture}
}
\caption{Token ring: Transducer-based approximation (2)}\label{fig:tr3}
\end{figure}
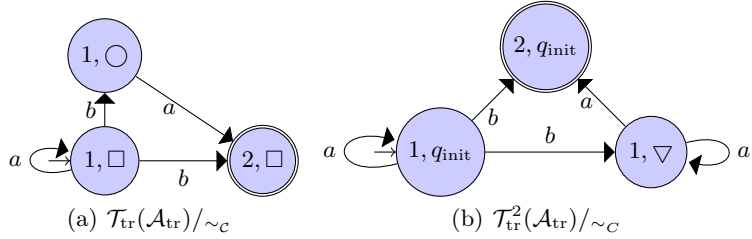

Let us consider again $\A_{\rm tr}$ and $\mathcal{T}_{\rm tr}$ from
Fig.~\ref{fig:tr1}. We consider the automaton $\C$ depicted in
Fig.~\ref{fig:tr2-a}. The automaton $\mathcal{T}_{\rm tr}^2(\A_{\rm
  tr})$ (after trimming) is depicted in Fig.~\ref{fig:tr2-b}. The
automata $\mathcal{T}_{\rm tr}(\A_{\rm tr})/_{\sim_\C}$ and
$\mathcal{T}_{\rm tr}^2(\A_{\rm tr})/_{\sim_\C}$ are depicted in
Fig.~\ref{fig:tr3}. For instance, in $\mathcal{T}_{\rm tr}^2(\A_{\rm
  tr})$ states $(1,3,4)$ and $(1,4,3)$ are $\sim_\C$-equivalent since
they have both $1$ as the first element, and $q_{\rm init}\cdot 34=q_{\rm
  init}\cdot 43=\bigtriangledown$.

\begin{figure}[tb]
\subfigure[$\C$]{\label{fig:tr2-a}
\begin{tikzpicture}[scale=0.7]
\node (init)[state,draw,fill=blue!20,initial, initial text=] at (0,0)
      {$q_{\rm init}$};
\node (rond)[state,draw,fill=blue!20,] at (2,1)
      {$\bigcirc$};

\node (carre)[state,draw,fill=blue!20,] at (2,-1)
      {$\square$};

\node (nabla)[state,draw,fill=blue!20,] at (4,0)
      {$\bigtriangledown$};

\path[->,>=triangle 90] (init) edge[] node [above] {$4$} (rond);
\path[->,>=triangle 90] (init) edge[bend left] node [above] {$3$} (carre);
\path[->,>=triangle 90] (carre) edge[bend left] node [above] {$3$} (init);
\path[->,>=triangle 90] (rond) edge[] node [above] {$3,4$} (nabla);
\path[->,>=triangle 90] (carre) edge[] node [above] {$4$} (nabla);
\path[->,>=triangle 90] (nabla) edge[loop above] node [above] {$3,4$} ();
\end{tikzpicture}
}\subfigure[$\mathcal{T}_{\rm tr}^2(\A_{\rm tr})$]{\label{fig:tr2-b}
\begin{tikzpicture}[scale=0.7]
\node (133)[state,draw,fill=blue!20,initial, initial text=] at (0,0) {$1,3,3$};

\node (233)[state,draw,fill=blue!20,accepting] at (3,0) {$2,3,3$};

\node (134)[state,draw,fill=blue!20,] at (6,1) {$1,3,4$};

\node (143)[state,draw,fill=blue!20,] at (6,-1) {$1,4,3$};

\path[->,>=triangle 90] (133) edge[loop above] node [right] {$a$} ();
\path[->,>=triangle 90] (133) edge[bend left] node [above] {$b$} (134);
\path[->,>=triangle 90] (133) edge[bend right] node [above] {$b$} (143);
\path[->,>=triangle 90] (133) edge[] node [above] {$b$} (233);

\path[->,>=triangle 90] (134) edge[] node [above] {$a$} (233);
\path[->,>=triangle 90] (134) edge[] node [left] {$a$} (143);

\path[->,>=triangle 90] (143) edge[] node [above] {$a$} (233);
\end{tikzpicture}
}
\caption{Token ring: Transducer-based approximation (1)}\label{fig:tr2}
\end{figure}

\begin{proposition}\label{prop-main}
An automaton isomorphic to $\T^k(\A)/_{\sim_\C}$ can be computed in
polynomial time in $k$ and in the sizes of $\A$, $\T$ and $\C$.
\end{proposition}

Now, given a finite automaton $\B$, we can use the computed automata
when applying the {\tt FixPointT} semi-algorithm described in
Fig.~\ref{algo:fpT}. It may provide an over-approximation of reachable
states: if {\tt FixPointT} stops on a not too coarse approximation we
can deduce that \hbox{$\R^*_\T(L(\A))\cap L(\B)=\emptyset$}. The proof
of Proposition~\ref{prop-FixedPointT} is similar to this of
Proposition~\ref{prop-fixed}.

\begin{proposition}\label{prop-FixedPointT}
The {\tt FixPointT} semi-algorithm  is correct: if it returns {\it safe}
then $\R^*_\T(L(\A))\cap L(\B)=\emptyset$.
\end{proposition}

\section{Refining Transducer-based Approximations}
\label{transducer-based-approximations-CEGAR}

In this section we propose to refine transducer-based approximations when the approximate iteration is inconclusive. 
Intuitively, this happens when the sequence of approximations is too coarse: the result intersects with the set of bad states after $k$ steps while the backward iteration of $k$ copies of the transducer from the bad states does not intersect with the initial states.
Our algorithm can be seen as a kind of CEGAR algorithms -- the paradigm introduced in~\cite{DBLP:conf/cav/ClarkeGJLV00} and intensively studied during the last decade (see for
example~\cite{bouajjani_abstract_2004,DBLP:conf/rta/BoichutCHK08}), with the aim of obtaining finer approximations/abstractions by exploiting counter-examples.

\begin{proposition}\label{prop-raff1}
If $L(\T^{k}_\C(\A))\cap L(\B)\neq\emptyset$, then either $L(\A)\cap
L(\T^{-k}(\B))\neq\emptyset$, or there exists $j$, $0\leq j\leq k$ such that
$L(\T^{j}_\C(\A))\cap L(\T^{j-k}(\B))\neq\emptyset$ and $L(\T(\T^{j-1}_\C(\A)))\cap
L(\T^{j-k}(\B))=\emptyset$.
\end{proposition}

\begin{figure}[tb]
\centering{
\begin{tikzpicture}[scale=0.9]

\node (A0text) at (-0.7,1.2) {{\footnotesize $L(\A)$}};
\node[draw, ellipse, text width=0.7cm, minimum height=2cm, fill=blue!20] (A0) at (0,0){};
\path (A0text) edge node {} (A0);

\node (A1text) at (1.3,1.5) {{\footnotesize $L(\T_\C(\A))$}};
\node (A1etext) at (1.3,-1.4) {{\footnotesize $L_1$}};
\node[draw, ellipse, text width=0.7cm, minimum height=2.4cm, fill=blue!20] (A1) at (2,0){};
\node[draw, ellipse, text width=0.5cm, minimum height=2.1cm,fill=blue!10] (A1e) at (2,0){};
\path (A1text) edge node {} (A1);
\path (A1etext) edge node {} (A1e);

\node (Aktext) at (4.3,1.7) {{\footnotesize $L(\T_\C^j(\A))$}};
\node (Aketext) at (3.7,-.6) {{\footnotesize $L_j$}};
\node[draw, ellipse, text width=0.7cm, minimum height=2.6cm, fill=blue!20] (Ak) at (5,0){};
\node[draw, ellipse, text width=0.5cm, minimum height=2.1cm,fill=blue!10] (Ake) at (5,0){};
\path (Aktext) edge node {} (Ak);
\path (Aketext) edge node {} (Ake);

\node (Ak+text) at (6.3,1.8) {{\footnotesize $L(\T_\C^{j+1}(\A))$}};
\node (Ak+etext) at (7.7,1.8) {{\footnotesize $L_{j+1}$}};
\node[draw, ellipse, text width=0.7cm, minimum height=2.8cm, fill=blue!20] (Ak+) at (7,0){};
\node[draw, ellipse, text width=0.5cm, minimum height=2.4cm,fill=blue!10] (Ak+e) at (7,0){};
\path (Ak+text) edge node {} (Ak+);
\path (Ak+etext) edge node {} (Ak+e);

\node (Aitext) at (9.3,1.8) {{\footnotesize $L(\T^{k-1}_\C(\A))$}};
\node (Aietext) at (8.5,-.8) {{\footnotesize $L_{k-1}$}};
\node[draw, ellipse, text width=0.7cm, minimum height=3cm, fill=blue!20] (Ai) at (10,0){};
\node[draw, ellipse, text width=0.5cm, minimum height=2.7cm,fill=blue!10] (Aie) at (10,0){};
\path (Aitext) edge node {} (Ai);
\path (Aietext) edge node {} (Aie);

\node (Ai+text) at (11,1.4) {{\footnotesize $L(\A_{k})$}};
\node (Ai+etext) at (11.3,2) {{\footnotesize $L_k$}};
\node[draw, ellipse, text width=0.7cm, minimum height=3.2cm, fill=blue!20] (Ai+) at (12,0){};
\node[draw, ellipse, text width=0.5cm, minimum height=2.9cm,fill=blue!10] (Ai+e) at (12,0){};
\path (Ai+text) edge node {} (Ai+);
\path (Ai+etext) edge node {} (Ai+e);

\path[->,>=triangle 60] (A0) edge[above,bend left] node
    {} (A1);
\path[->,>=triangle 60] (A1) edge[above,bend left,dashed] node
    {} (Ak);
\path[->,>=triangle 60] (Ak) edge[above,bend left] node
    {} (Ak+);
\path[->,>=triangle 60] (Ai) edge[above,bend left] node
    {} (Ai+);

\path[->,>=triangle 60] (Ak+) edge[above,bend left,dashed] node
    {} (Ai);

\node (Aproptext) at (11.8,-3.3) {{\footnotesize $L(\B)$}};
\node[draw, ellipse, text width=0.5cm, minimum height=1.2cm, fill=red!20, opacity=0.8] (Aprop) at (12,-2){};
\path (Aproptext) edge node {} (Aprop);

\node (A1proptext) at (8.8,-3.3) {{\footnotesize $L(\T^{-1}(\B))$}};
\node[draw, ellipse, text width=0.5cm, minimum height=1.4cm, fill=red!20, opacity=0.8] (A1prop) at (10,-2){};
\path (A1proptext) edge node {} (A1prop);

\node (Ai+proptext) at (5.8,-3.3) {{\footnotesize $L(\T^{j+1-k}(\B))$}};
\node[draw, ellipse, text width=0.5cm, minimum height=1.6cm, fill=red!20, opacity=0.8] (Ai+prop) at (7,-2){};
\path (Ai+proptext) edge node {} (Ai+prop);

\node (Aiproptext) at (3,-2.8) {{\footnotesize $L(\T^{j-k}(\B))$}};
\node[draw, ellipse, text width=0.5cm, minimum height=1.8cm, fill=red!20, opacity=0.8] (Aiprop) at (4.4,-2){};
\path (Aiproptext) edge node {} (Aiprop);

\path[->,>=triangle 60] (Aprop) edge[above,bend right] node
    {} (A1prop);

\path[->,>=triangle 60] (A1prop) edge[above,bend right,dashed] node
    {} (Ai+prop);

\end{tikzpicture}
}
\caption{Refinement: $L_i$'s represent the languages 
$L(\T(\T^{i-1}_\C(\A))$'s.}\label{fig-raffinement}
\end{figure}
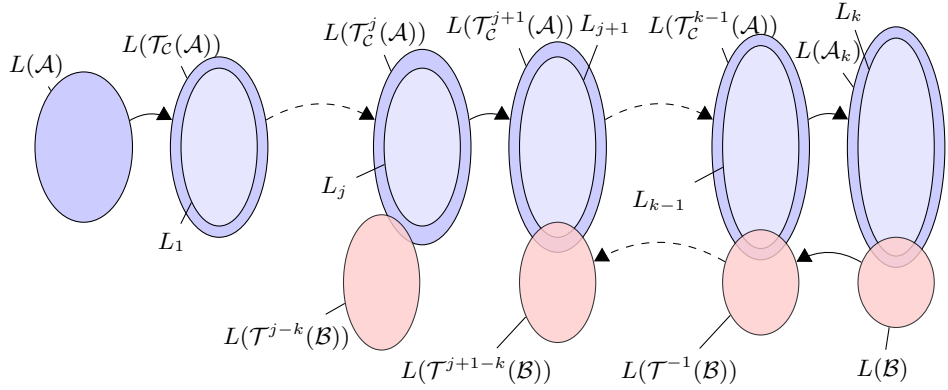




Assume that $L(\T^{j}_\C(\A))\cap L(\T^{j-k}(\B))\neq\emptyset$ and
$L(\T(\T^{j-1}_\C(\A)))\cap L(\T^{j-k}(\B))=\emptyset$. As it is
classically done in the CEGAR framework, one can compute a relation
$\equiv$ on $\T^{j}_\C(\A)$ such that $\equiv\subseteq \sim_{\C}$ and $L(\T^{j}_\C(\A))/_\equiv\cap
L(\T^{k-j}(\B))=\emptyset$. The existence of $\equiv$ is trivial since
the results hold for the identity relation.  However, when using the
CEGAR approach, our goal is to compute a relation $\equiv$ as large as
possible, with the aim of ensuring termination of state-space exploration. 

To achieve this goal, several heuristics may be used.  Instead of
computing the $\equiv$ relation, building the corresponding
$\T^{j}_\C(\A)/_\equiv$ automaton, and then performing the fixpoint
computation, we propose to use a dynamic approach.  More precisely, we
prefer to modify $\C$ according to $\equiv$ to avoid similar states
merging which may lead to a coarser over-approximation. To modify $\C$
according to $\equiv$, we propose to use the algorithms in
Figs.~\ref{algo:raff} and~\ref{algo:raff2}. The {\tt Split} algorithm
modifies the given deterministic automaton to provide a weaker
abstraction.  Its idea is quite natural: if two equivalent states must
be distinguished, the automaton $\C$ is refined to take this
constraint into account.  For example, Figure~\ref{fig:split-a}
displays the automaton $\C'$ resulting from
$\mathtt{Split}(\C,\bigcirc,\square,3,4)$, where $\C$ is the automaton
from Fig.~\ref{fig:tr2-a}. %
The {\tt Split} algorithm dissociating two states, can be used so far
as necessary to obtain the refined approximation in the {\tt Refine}
algorithm in Fig.~\ref{algo:raff2}.

\begin{figure}[tb]
{\bf Algorithm} {\tt Split}\\
{\bf Input:} $S=(Q_S,Q_T,E_S,\{q_0\},\emptyset)$ a deterministic automaton, $p,q\in
Q_S$ and $\alpha,\beta\in Q_T$ such that $p\cdot_S \alpha=q\cdot_S \beta$\\
\spt $Q_S^\prime:=Q_S\cup\{r\}$ where $r\notin Q_S$\\
\spt $I^\prime_S:=\{q_0\}$\\
\spt $E_S^\prime:=E_S\setminus \{(q,\beta,q\cdot_S\beta)\}$\\
\spt $E_S^\prime:=E_S^\prime\cup
\{(q,\beta,r)\}\cup\{(r,a,s)\mid (p\cdot\alpha,a,s)\in
E_S \text{ and } s\in Q_S\setminus\{p\cdot_S\alpha\}\}$\\
\spt $E_S^\prime:=E_S^\prime\cup\{(r,a,r)\mid (p\cdot\alpha,a,p\cdot\alpha)\in
E_S \}$\\
\spt {\bf Return} $(Q^\prime_S,Q_T,E_S^\prime,I_S^\prime,\emptyset)$
\caption{Algorithm {\tt Split}}\label{algo:raff}
\end{figure}

\begin{figure}[tb]
\subfigure[$\C^\prime=\mathtt{Split}(\C,\bigcirc,\square,3,4)$]{\label{fig:split-a}
\begin{tikzpicture}[scale=0.7]
\node (init)[state,draw,fill=blue!20,initial, initial text=] at (0,0)
      {$q_{\rm init}$};
\node (rond)[state,draw,fill=blue!20,] at (2,1)
      {$\bigcirc$};

\node (carre)[state,draw,fill=blue!20,] at (2,-1)
      {$\square$};

\node (nabla)[state,draw,fill=blue!20,] at (4,1)
      {$\bigtriangledown$};
\node (r2)[state,draw,fill=blue!20,] at (4,-1)
      {$r$};

\path[->,>=triangle 90] (init) edge[] node [above] {$4$} (rond);
\path[->,>=triangle 90] (init) edge[bend left] node [above] {$3$} (carre);
\path[->,>=triangle 90] (carre) edge[bend left] node [above] {$3$} (init);
\path[->,>=triangle 90] (rond) edge[] node [above] {$3,4$} (nabla);
\path[->,>=triangle 90] (carre) edge[] node [above] {$4$} (r2);
\path[->,>=triangle 90] (nabla) edge[loop above] node [above] {$3,4$} ();
\path[->,>=triangle 90] (r2) edge[loop right] node [above] {$3,4$} ();
\end{tikzpicture}
}\subfigure[$\T_{\rm tr}(\T_{\rm tr}(\A_{\rm
  tr})/_{\sim_\C})/_{\sim_{\C^\prime}}$]{\label{fig:split-b}
\begin{tikzpicture}[scale=0.7]
\node (133)[state,draw,fill=blue!20,initial, initial text=] at (0,-1)
      {$1,q_{\rm init}$};
\node (143)[state,draw,fill=blue!20,] at (-2,1)
      {$1,r_1$};

\node (233)[state,draw,fill=blue!20,accepting] at (2,1)
      {$2,q_{\rm init}$};

\node (134)[state,draw,fill=blue!20,] at (4,-1)
      {$1,r_2$};


\path[->,>=triangle 90] (133) edge[loop left] node [above] {$a$} ();
\path[->,>=triangle 90] (133) edge[] node [left] {$b$} (143);
\path[->,>=triangle 90] (133) edge[] node [above] {$b$} (134);
\path[->,>=triangle 90] (133) edge[] node [below] {$b$} (233);
\path[->,>=triangle 90] (143) edge[] node [above] {$a$} (233);
\path[->,>=triangle 90] (134) edge[] node [left] {$a$} (233);
\path[->,>=triangle 90] (134) edge[] node [left] {$a$} (143);
\path[->,>=triangle 90] (143) edge[] node [above] {$a$} (233);
\end{tikzpicture}
}
\caption{Examples for the {\tt Split} and {\tt Refine} algorithms}\label{fig:split}
\end{figure}
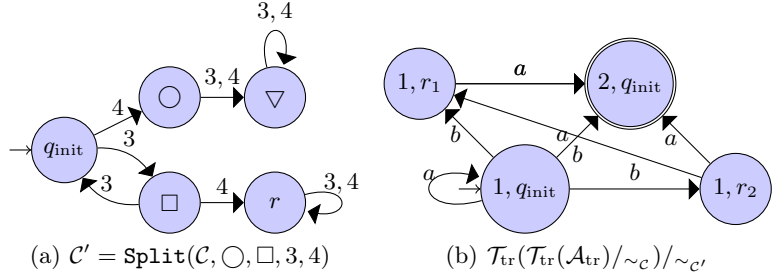

\begin{figure}[tb]
{\bf Algorithm} {\tt Refine}\\
{\bf Input:} $\T$ (transducer), $\C$ a deterministic automaton,
$S=(Q_S\times Q_C,Q,E,\{q_0\},F_S)$ a finite automaton, a relation $\equiv$
such that $\equiv\subseteq \sim_{\C}$ and 
$L(\T_\C(\A))/_\equiv\cap L(\T^{-1}(\B))=\emptyset$\\
\sp {\bf While} ($\sim_\C\not\subseteq\equiv$) {\bf do}\\
\sp\sp {\bf Choose} $(p,q,\alpha)$ {\bf and} $(p,q^\prime,\alpha^\prime)$
states of $\T(S)$ {\bf such that}\\
\sp\sp\sp$(p,q,\alpha)\sim_\C(p,q^\prime,\alpha^\prime)$ but
$(p,q,\alpha)\not\equiv(p,q^\prime,\alpha^\prime)$ \\
\sp\sp $\C:=${\tt Split}$(\C,q,\alpha,q^\prime,\alpha^\prime)$ \\
\sp {\bf EndWhile}\\ 
\sp {\bf Return} $\C$
\caption{Algorithm {\tt Refine}}\label{algo:raff2}
\end{figure}

\begin{proposition}\label{prop-refine}
  The  {\tt Refine} algorithm always terminates.
\end{proposition}

For example, let us consider the $\equiv$ relation whose classes are
$\{1,\square,4\}$, $\{(2,\square,3)\}$, $\{(1,\bigcirc,3),(1,\bigcirc,4)\}$
and $\{(1,\square,3)\}$.  We apply the {\tt
  Refine} algorithm to the automata $\T_{\rm tr}$
(Fig.~\ref{fig:tr-T}), $\C$ (Fig.~\ref{fig:tr2-a}), $\T_{\rm
  tr}(\A_{\rm tr})/_{\sim_\C}$ (Fig.~\ref{fig:tr3-a}).  Since
$(1,\bigcirc,3)\sim_\C (1,\square,4)$, $\sim_\C\not\subseteq
\equiv$. Therefore, the algorithm may compute
$\C^\prime=\mathtt{Split}(\C,\bigcirc,\square,3,4)$ as depicted in
Fig.~\ref{fig:split-a}. Then one can check that
$\sim_{\C^\prime}\subseteq \equiv$. The automaton $\T_{\rm tr}(\T_{\rm
  tr}(\A_{\rm tr})/_{\sim_\C})/_{\sim_{\C^\prime}}$ is depicted in
Fig.~\ref{fig:split-b}.

\begin{figure}[tb]
{\bf Semi-Algorithm} {\tt Reach-CEGAR}\\
{\bf Input:} $\A,\B$  finite automata, $\T$ (transducer), $\C$ a
deterministic automaton, an integer~$\ell$\\
{\bf Variables:} integers $j,k$, and equivalence relation $\equiv$\\
\sp $k:=\ell$\\
\sp {\bf While} ($L(\T_\C^k(\A))\cap L(\B)= \emptyset$ {\bf and}
$L(\T_\C^{k+1}(\A))\neq L(\T_\C^k(\A))$ ) {\bf do}\\
\sp\sp $k:=k+1$\\
\sp {\bf EndWhile}\\
\sp {\bf If} ($L(\T_\C^{k+1}(\A))=L(\T_\C^k(\A))$ {\bf and} $L(\T_\C^k(\A))\cap L(\B)= \emptyset$ ) {\bf then}\\
\sp\sp {\bf Return} {\it Safe}\\
\sp {\bf EndIf}\\
\sp {\bf If}  $L(\A)\cap L(\T^{-k}(\B)) \neq \emptyset$ {\bf then}\\
\sp\sp {\bf Return} {\it Unsafe}\\
\sp{\bf EndIf}\\
\sp $j:=\mathtt{J}(\A,\B,\C,\T,k)$\\
\sp {\bf Let} $\equiv$ be such that $\equiv\subseteq \sim_{\C}$ and 
$L(\T^{j}_\C(\A))/_\equiv\cap L(\T^{k-j}(\B))=\emptyset$\\
\sp {\bf Return}  {\tt Reach-CEGAR}$(\A,\T^{-k}(\B),\T,\mathtt{Refine}(\T,\C,\T^j(\A),\equiv),j)$
\caption{Semi-algorithm {\tt Reach-CEGAR}}\label{algo:cegar}
\end{figure}

If $L(\T^{k}_\C(\A))\cap L(\B)\neq\emptyset$ and $L(\A)\cap
L(\T^{-k}(\B))=\emptyset$, then we denote by
$\mathtt{J}(\A,\B,\C,\T,k)$ the maximal integer $j$ such that $0\leq
j\leq k$ and $L(\T^{j}_\C(\A))\cap L(\T^{j-k}(\B))\neq\emptyset$ and
$L(\T(\T^{j-1}_\C(\A)))\cap L(\T^{j-k}(\B))=\emptyset$. Now, the {\tt
  Reach-CEGAR} semi-algorithm in Fig.~\ref{algo:cegar}
encodes the whole approach: each time a too strong approximation is
detected, it is refined. This semi-algorithm may terminate by
returning {\it Safe} if an over-approximation of accessible states
that does not contain any bad states. It may also
terminate by returning {\it Unsafe} if it detects a reachable bad state. 
It may also diverge if the computed approximations have to
be refined again and again.


\section{Experimental Results}\label{sec:experimentations}

Thanks to a prototype tool, the present paper's proposals have been evaluated on the
well-known examples of the
Bakery algorithm by Lamport, the token ring algorithm, Dijkstra's, and
Burns~\cite{touili_regular_2001} protocols.


For the quotient-based approximations
(Sect.~\ref{quotient-based-approximations}), the results are displayed in
Fig.~\ref{tableau_resultats_master}. 
In addition to $\mathfrak{Left}$ and $\mathfrak{Right}$, 
two additional simple isomorphism-compatible approximations are examined: \\
\noindent -- $\mathfrak{In}$, mapping each automaton $(Q,\Sigma,E,I,F)$ to the
   reflexive-transitive closure of the relation $R_{\rm in}$, defined by $p R_{\rm in} q$ iff
 $\lbrace a_p \in \Sigma \mid \exists p' \in Q, (p',a_p,p) \in E \rbrace = \lbrace a_q \in \Sigma \mid \exists q' \in Q, (q',a_q,q) \in E \rbrace$; and \\
%
\noindent -- $\mathfrak{Out}$, mapping each automaton $(Q,\Sigma,E,I,F)$ to the
   reflexive-transitive closure of the relation $R_{\rm out}$, defined by $p R_{\rm out} q$ iff
 $\lbrace a_p \in \Sigma \mid \exists p' \in Q, (p,a_p,p') \in E \rbrace = \lbrace a_q \in \Sigma \mid \exists q' \in Q, (q,a_q,q') \in E \rbrace$.\\
 In Fig.~\ref{tableau_resultats_master}, the first column describes
 the protocol to verify: its name, the size (i.e., $|Q|+|E|$) of the
 initial automaton, and that of the transducer. The remaining columns give the results for each specific criterion:
the first line gives the step of the language equality, or No when not reached;
the second line indicates the step when the intersection with the bad-property language is non empty, or $\emptyset$ if it remains empty;
the third line gives the size of the last obtained automaton.
If a step of the languages equality occurs while having the empty intersection with the bad-property language (cf. values highlighted in bold), the protocol is safe.


For the refinement method, the above mentioned protocols have been studied using
different kinds of $\C$-automata: either a {\it one-state} $\C$, or a
{\it specific} $\C$.  When starting the refinement with a one-state
$\C$ in Fig.~\ref{fig:one-state-b}, all the states are obviously
considered as $\C$-equivalent. On the contrary, a specific $\C$ models
a property of interest. For example, if two consecutive $a$ are
forbidden, and there is a transition $(p, (x, a), q)$ in the
transducer of the considered protocol, then the
specific $\C$ is like in Fig.~\ref{fig:specific-B}.  The two token
ring protocols are shown to be safe  in four
steps using the refinement approach with a one-state automaton. Dijkstra's protocol was proved
safe without refinement in 15 steps using a specific automaton. The
Bakery and Burns protocols are proved safe  in respectively 6 and 14 steps, by 
using the refinement and specific automata. For all these
protocols, the obtained automata have sizes similar to the sizes of
the input automata: there is no state explosion. To conclude, the experiments show that our techniques work for all the
considered cases,  and that they are complementary. 


\begin{figure}[tb]
\hfill
\subfigure[one-state $\C$]{\label{fig:one-state-b}
\begin{tikzpicture}[scale=0.7]
\node (i)[state,draw,fill=blue!20,initial, initial text=] at (0,0) {$i$};
\path[->,>=triangle 90] (i) edge[loop above] node [above] {$\Sigma$} ();
\end{tikzpicture}
}\hfill
\subfigure[specific $\C$ for mutual exclusion protocols]{\label{fig:specific-B}
\begin{tikzpicture}[scale=0.7]
\node (i)[state,draw,fill=blue!20,initial, initial text=] at (0,0) {$i$};
\node (bp)[state,draw,fill=blue!20] at (3,0.9) {$b_p$};
\node (bq)[state,draw,fill=blue!20] at (3,-0.9) {$b_q$};
\path[->,>=triangle 90] (i) edge[out=45] node [below] {$p$} (bp);
\path[->,>=triangle 90] (i) edge[out=-45, in=225] node [above] {$q$} (bq);
\path[->,>=triangle 90] (i) edge[loop above] node [above] {$\Sigma - \{p,q\}$} ();

\path[->,>=triangle 90] (bp) edge[out=225, in=135] node [left] {$q$} (bq);
\path[->,>=triangle 90] (bp) edge[out=10, in=40, looseness=7] node [right] {$\Sigma - \{q\}$} (bp);

\path[->,>=triangle 90] (bq) edge[out=45, in=-45] node [right] {$p$} (bp);
\path[->,>=triangle 90] (bq) edge[out=-10, in=-40, looseness=7] node [right] {$\Sigma - \{p\}$} (bq);
\end{tikzpicture}
}
\hfill
\caption{Different kinds of $\C$ automata}
\end{figure}
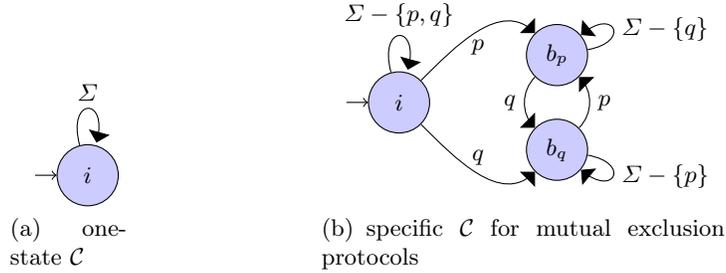

\begin{figure}[tb]
\begin{bigcenter}
\begin{scriptsize}
\begin{tabular}{|c|c|c|c|c|c|c|c|c|c|c|}
\hline
			&	$\mathfrak{In}$		&	$\mathfrak{Out}$		&	$\mathfrak{In+Out}$		&	$\mathfrak{In.Out}$		&	$\mathfrak{Left}$			&	$\mathfrak{Right}$			&	$\mathfrak{L+R}$			&	$\mathfrak{L.R}$			&	$\mathfrak{(L+R).(In+Out)}$	\\
\hline
Token ring		&	\textbf{Step 3}			&	\textbf{Step 3}			&	\textbf{Step 3}			&	\textbf{Step 4}			&	\textbf{Step 3}	&	\textbf{Step 2}	&	\textbf{Step 2}	&	\textbf{Step 3}	&	\textbf{Step 3}	\\
size I : 4		&	$\mathbf{\emptyset}$	&	$\mathbf{\emptyset}$	&	$\mathbf{\emptyset}$	& 	$\mathbf{\emptyset}$	&	$\mathbf{\emptyset}$		&	$\mathbf{\emptyset}$		&	$\mathbf{\emptyset}$		&	$\mathbf{\emptyset}$		&	$\mathbf{\emptyset}$		\\
size T : 6		&	8					&	8					&	5					&	12					&	8				&	5				&	5				&	8				&	5				\\
\hline
Token ring		&	\textbf{Step 3}			&	\textbf{Step 3}			&	\textbf{Step 3}			&	\textbf{Step 4}			&	T.o(Step 10)		&	\textbf{Step 2}	&	\textbf{Step 2}	&	T.o(Step 10)		&	\textbf{Step 3}	\\
size I : 4		&	$\mathbf{\emptyset}$	&	$\mathbf{\emptyset}$	&	$\mathbf{\emptyset}$	& 	$\mathbf{\emptyset}$	&	$\emptyset$				&	$\mathbf{\emptyset}$		&	$\mathbf{\emptyset}$		&	$\emptyset$				&	$\mathbf{\emptyset}$		\\
size T : 9		&	8					&	8					&	5					&	12					&	109				&	5				&	5				&	109				&	5				\\
\hline
Dijkstra		&	\textbf{Step 6}			&	\textbf{Step 6}			&	\textbf{Step 5}			&	\textbf{Step 7}			&	T.o(Step 10)		&	\textbf{Step 5}	&	\textbf{Step 5}	&	T.o(115 hours)		&	\textbf{Step 5}	\\
size I : 5		&	$\mathbf{\emptyset}$	&	$\mathbf{\emptyset}$	&	$\mathbf{\emptyset}$	&	$\mathbf{\emptyset}$	&	$\emptyset$				&	$\mathbf{\emptyset}$		&	$\mathbf{\emptyset}$		&	T.o(115 hours)		&	$\mathbf{\emptyset}$		\\	
size T : 62		&	49					&	118					&	20					&	246					&	745				&	11				&	10				&	T.o(115 hours)		&	15				\\	
\hline
Bakery		&	\textbf{Step 7}			&	No					&	No					&	\textbf{Step 10}			&	T.o(Step 10)		&	No				&	No				&	T.o(Step 10)		&	No				\\
size I : 2		&	$\mathbf{\emptyset}$	&	Step 7				&	Step 6				&	$\mathbf{\emptyset}$	&	$\emptyset$				&	Step 3			&	Step 3			&	$\emptyset$				&	Step 6			\\
size T : 24		&	43					&	75					&	89					&	97					&	368				&	31				&	31				&	1253			&	101				\\	
\hline
Burns		&	No					&	\textbf{Step 6}			&	No					&	No					&	No				&	No				&	No				&	No				&	No				\\
size I : 2		&	Step 5				&	$\mathbf{\emptyset}$	&	Step 3				&	Step 7				&	Step 4			&	Step 3			&	Step 3			&	Step 4			&	Step 3			\\	
size T : 22		&	100					&	46					&	53					&	365					&	22				&	18				&	18				&	50				&	53				\\	
\hline
\end{tabular}
\end{scriptsize}
\caption{Results with syntactic criteria}
\label{tableau_resultats_master}
\end{bigcenter}
\end{figure}

\section{Conclusion}

Developing efficient approximation-based techniques is a critical
challenging issue to tackle reachability problems when exact
approaches 
do not work.  In this paper
two new approximation techniques for the regular reachability problem
have been presented. 
Our techniques use polynomial time algorithms, provided that 
recent algorithms for checking automata equivalence are used; the only exception being language inclusion testing 
as in~\cite{DBLP:conf/tacas/DoyenR10,DBLP:conf/tacas/AbdullaCHMV10,BONCHI-2012-674660}.
As a future direction, we plan to upgrade our refinement approach, 
both on the precision of the
approximations and on computation time. Another possible direction is to generalize 
our approximation mechanisms and to apply them to 
other RMC applications, e.g.,  
counter systems or push-down systems.

\bibliographystyle{plain}
\bibliography{biblio}

\end{document}